

\magnification=\magstep1
\font\fontina=cmr8
\baselineskip=15pt

\def\p{\partial}
\def\py{\partial y}

\def\tr{{\rm tr}}
\def\det{{\rm det}}

\overfullrule=0pt

\line{\hfil CTP-TAMU-15/94}
\line{\hfil SISSA 44/94/EP}
\line{\hfil hepth@xxx/9407021}
\line{\hfil June 1994}
\vskip 2.4truecm
\centerline{\bf On Target-Space Duality in $p$--Branes}
\vskip 1truecm
\centerline{ R. Percacci}
\bigskip
\centerline{\it International  School for Advanced Studies,
via Beirut 4, 34014 Trieste, Italy}
\centerline{\it and Istituto Nazionale di Fisica Nucleare, Sezione di
Trieste.}
\bigskip
\centerline{and}
\bigskip
\centerline{E. Sezgin \footnote{$^\ast$}{\fontina Supported in part by the
U.S.\ National Science Foundation, under grant PHY-9106593.}}
\bigskip
\centerline{\it Center for Theoretical Physics, Texas A\&M University,}
\centerline{\it College Station, Texas 77843-4242, U.S.A.}
\vskip 1.4truecm
\centerline{\bf ABSTRACT}
\smallskip
\midinsert\narrower{
We study the target-space duality transformations in $p$--branes as
transformations which mix the worldvolume field equations with Bianchi
identities. We consider an $(m+p+1)$-dimensional spacetime with
$p+1$ dimensions compactified, and a particular form of the
background  fields.
We find that while a $GL(2)=SL(2)\times R$ group is realized
when $m=0$, only a two parameter group is realized when $m>0$.
}\endinsert \vfill\eject
\medskip

Ten dimensional string theory compactified on a six--torus has a
symmetry group $O(6,6;Z)$, called ``T duality'', which mixes momentum
modes with winding modes [1-4] (in the heterotic case the group is
$O(6,22;Z)$). This is also referred to as
{\it target\ space\ duality.} (For an extensive review and further references,
see [5]). It has been conjectured that string theory possesses also a symmetry
group $SL(2,Z)$, called ``S duality'', which transforms Fock states
into soliton states [6-8]. This is also known as
{\it strong--weak\ coupling\ duality.} (For a recent review, see [9]).
Such a symmetry could only be seen in a
nonperturbative approach and its existence has not yet been firmly
established.

One possible way of studying the S duality of strings is via
fivebranes. A fivebrane in ten dimensions has been conjectured to be
``dual'' to a string in a certain well defined sense [10],
and it has been further conjectured that under this transformation
the role of the S and T dualities would interchange [11], in the sense that
the $SL(2,Z)$ S duality of the string would play the role of a T duality
for the fivebrane, and similarly the $O(6,6;Z)$ T duality of the
string would play the role of a nonperturbative S duality for the
fivebrane.
Since a T duality symmetry can be seen perturbatively, it would seem
convenient to study $SL(2,Z)$ symmetry in the context of five--branes.

All these groups have been found either directly in four dimensional
$N=4$ supergravity [3] or in the toroidal compactification of ten dimensional
supergravity
theories [4] which one obtains as low energy limits of strings and
fivebranes.
It was conjectured by Duff and proven by Cecotti et al. that
the T duality group $O(6,6;Z)$ is a symmetry of string theory,
transforming the
worldsheet field equations and Bianchi identities into each other [2].
Although it has been shown that an $SL(2,Z)$ duality group indeed mixes the
momentum modes with the winding modes of the five-brane theory [7], so far
this symmetry group has not been understood as a worldvolume duality group
transforming the worldvolume equations of motion to Bianchi identities.

This is the problem that we will investigate in the present paper.
Although the case of greatest interest is the five-brane in $4+6$
dimensions, of which $6$ are compact, for the sake of generality
we shall discuss arbitrary $p$-branes in $m+p+1$ dimensions,
of which $p+1$ are compact.
We will assume a specific simple form of the worldvolume action [12].
In the Lagrangian formulation, we show that while an
$SL(2)$ symmetry is realized for $m=0$,
only a two parameter algebra formed by a particular set of upper triangular
matrices is realized for $m>0$.

An attempt to find duality symmetries in $p$-branes has been made
before, in a somewhat different setting [13].
We shall comment on some results of this paper in the end.

The duality symmetries of (compactified) strings and the conjectured
duality symmetries of five-branes form discrete groups.
We assume that at a local level the duality symmetries actually form
continuous groups, and that they are broken down to discrete groups
by the boundary conditions due to the compactification. This is known
to be the case for the string.
Therefore, in discussing the duality symmetry of the equations of
motion, we can restrict ourselves to infinitesimal transformations.
The duality group is determined by the
consistency requirements that were first discussed in this context
by Gaillard and Zumino [14].

Before considering the case of general $p$--branes, let us outline
how these requirements work in the simpler model
of a string on a target space of dimension $m+n$, with $n$ dimensions
compactified. The dynamical variables are
$x^\mu(\sigma)$, $y^\alpha(\sigma)$ and a world-sheet metric
$\gamma_{ij}(\sigma)$.
Here $\sigma^i$  $(i=0,1)$ are the world-sheet coordinates,
$x^\mu$, $\mu=0,...,m-1$ are coordinates on $m$ dimensional spacetime $M$ and
$y^\alpha$, $\alpha=1,...,n$ are coordinates on an internal
$n$--dimensional manifold $N$.
For simplicity, we assume that the only nonvanishing background fields
are the metrics $g_{\mu\nu}(x)$ and $g_{\alpha\beta}(x)$ on $M$ and $N$
respectively  and an antisymmetric tensor field $b_{\alpha\beta}(x)$.
Although $N$ is supposed to be compact, we shall mostly not be concerned
with the boundary conditions that are implied by this fact.
At the local level at which we shall work the only distinction between the
$y^\alpha$ and the $x^\mu$ coordinates is that the antisymmetric tensor $b$
has vanishing $x^\mu$ components.

The action for the string is
$$
\eqalign{
S =\int d^2\sigma {\cal L}
= \int d^2\sigma \Bigl[-{1\over2}\sqrt{-\gamma}
\bigl(\gamma^{ij}\partial_i x^\mu\partial_j x^\nu g_{\mu\nu} &
+\gamma^{ij}\partial_i y^\alpha \partial_j y^\beta g_{\alpha\beta}\bigr)
\cr
&+{1\over2} \epsilon^{ij}\partial_i y^\alpha \partial_j y^\beta
b_{\alpha\beta}\Bigr]\ .\cr}\eqno(1)
$$

The field equations for the variables $x^\mu$ following from the action (1)
are
$$
g_{\mu\nu}\p_i\bigl(\sqrt{-\gamma}\gamma^{ij}\p_j x^\nu\bigr)
+\sqrt{-\gamma}\gamma^{ij}\p_i x^\rho\p_j
x^\nu{\mit\Gamma}_{\mu,\rho\nu}
+{1\over 2}\left({\gamma_{ij}\over\sqrt{-\gamma}}\p_\mu g_{\alpha\beta}
-\epsilon_{ij}\p_\mu b_{\alpha\beta}\right)
J^{i\alpha}J^{j\beta}=0\ , \eqno(2)
$$
where ${\mit\Gamma}_{\mu,\rho\nu}$ are the Christoffel symbols of the
first kind for the metric $g_{\mu\nu}$ and we have defined the conserved
current
\footnote{$^\ast$}{In our conventions the world-sheet Levi--Civita symbols
are defined by $\epsilon^{01}=-\epsilon_{01}=1$,
the world-sheet signature is $(-+)$ and the spacetime signature
$(-+\cdots +)$. The Levi--Civita symbols on $N$ have components
$\epsilon^{12\ldots n}=\epsilon_{12\ldots n}=1$}
$$
J^{i\alpha}=\epsilon^{ij} \partial_j y^\alpha \ . \eqno(3)
$$
The field equations for the variables $y^\alpha$ can be
written in the form
$$ \partial_i P^i{}_\alpha=0\ , \eqno(4)
$$
where
$$
P^i{}_\alpha={\partial{\cal L}\over\partial\partial_i y^\alpha}=
\left(-\sqrt{-\gamma}\gamma^{ij}g_{\alpha\beta}
+\epsilon^{ij}b_{\alpha\beta}\right)\epsilon_{jk} J^{k\beta}\ ,
\eqno(5)
$$
Finally, the equations of motion of the metric $\gamma$ state that the
energy--momentum tensor of $x^\mu$ and $y^\alpha$, regarded as scalar
fields on the world-sheet, has to be zero.
{}From the definition (3) follows the identity
$$
\p_i J^{i\alpha}=0\ . \eqno(6)
$$
This equation will be referred to as a ``Bianchi identity''. The duality
transformations we seek for mix the
field equation (4) and Bianchi identitity (6). To this end let us consider
the following infinitesimal transformations
$$
\eqalign{
\delta P^i{}_\alpha =& A_\alpha{}^\beta P^i{}_\beta
      +B_{\alpha\beta}J^{i\beta}\ , \cr
\delta J^{i\alpha} =& C^{\alpha\beta}P^i{}_\beta+D^\alpha{}_\beta
J^{i\beta}\ , \cr} \eqno(7)
$$
where $A,B,C,D$ are constant matrices. We now have to check that these
transformations together with appropriate transformation rules for
the background fields, which are to be determined, actually leave
invariant Eq. (5). Varying (5) we obtain
$$
  \eqalign{
A_\alpha{}^\beta P^i{}_\beta
      +B_{\alpha\beta}J^{i\beta} &=
\left(-\sqrt{-\gamma}\gamma^{ij}\delta g_{\alpha\beta}
+\epsilon^{ij}\delta b_{\alpha\beta}\right)\epsilon_{jk} J^{k\beta}\cr
&+\left(-\sqrt{-\gamma}\gamma^{ij}g_{\alpha\beta}
+\epsilon^{ij} b_{\alpha\beta}\right)\epsilon_{jk}
\left( C^{\beta\gamma}P^k{}_\gamma+D^\beta{}_\gamma J^{k\gamma}\right)\ .\cr}
\eqno(8)
$$
Eliminating $P^i{}_\alpha$ by using (5), we express all the terms in
(8) in terms of $J^i{}_\alpha$. Thus, demanding that (8) is satisfied we
arrive at the conditions $B_{\alpha\beta}=-B_{\beta\alpha}$,
$C^{\alpha\beta}=-C^{\beta\alpha}$,
$D^\alpha{}_\beta=-A_\beta{}^\alpha$
and the background field transformation rules
$$
\eqalign{
\delta g_{\alpha\beta}=& A_\alpha{}^\gamma g_{\gamma\beta}+
A_\beta{}^\gamma g_{\gamma\alpha}
-g_{\alpha\gamma}C^{\gamma\delta}b_{\delta\beta}
-g_{\beta\gamma}C^{\gamma\delta}b_{\delta\alpha}\ , \cr
\delta b_{\alpha\beta}=& A_\alpha{}^\gamma b_{\gamma\beta}-A_\beta{}^\gamma
b_{\gamma\alpha} -g_{\alpha\gamma}C^{\gamma\delta}g_{\delta\beta}
-b_{\alpha\gamma}C^{\gamma\delta}b_{\delta\beta}+ B_{\alpha\beta} \ .\cr}
\eqno(9)
$$
At this point one can check that equation (2) as well as the equation
of motion for $\gamma$ are also invariant under this set of transformations.
It is important to observe that in the case of the string one can consistently
take $\gamma$ to be inert under duality transformations.
In fact, the equation of motion for $\gamma$ has the general solution
$\gamma_{ij}=\Omega^2\left(\partial_i x^\mu\partial_j x^\nu g_{\mu\nu}
+\partial_i y^\alpha\partial_j y^\beta g_{\alpha\beta}\right)$,
where $\Omega$ is an undetermined scale factor.
Using equations (3), (7) and (9) one can show that $\gamma_{ij}$
is indeed duality invariant.

The conditions on $A,B,C,D$ imply that they form the group $O(n,n;R)$.
(In the heterotic case one would find $O(n_1,n_2;R)$, with $n_1$ and $n_2$
referring to the number of internal left and right movers).
All this was at the local level. Taking into account the boundary
conditions due to the compactness of $N$, one finds the
duality group $O(n,n;Z)$.

In this paper we will be interested in the case of $p$-branes
when the number of compact dimensions is $p+1$.
In order to extract some more information from the
string case, let us therefore see what happens in the case of the string
when $n=2$.
Apart from reflections, the group $O(2,2;R)$ is the direct product
of two commuting $SL(2)$ subgroups. The Lie algebra of
one of these groups consists of
block diagonal matrices with $B=C=0$, $D=-A^T$ and $A$ traceless.
{}From (7) we see that it does not mix the field equations with the Bianchi
identities, and therefore we shall refer to it as the ``trivial'' $SL(2)$.
The other $SL(2)$ is defined by
$A_\alpha{}^\beta=a\delta_\alpha^\beta$,
$B_{\alpha\beta}=b\epsilon_{\alpha\beta}$,
$C^{\alpha\beta}=-c\epsilon^{\alpha\beta}$
and $D^\alpha{}_\beta=-a\delta^\alpha_\beta$.

Let us now consider background fields of the special form
$g_{\alpha\beta}=\lambda_2 \delta_{\alpha\beta}$,
$b_{\alpha\beta}=\lambda_1 \epsilon_{\alpha\beta}$,
where $\lambda_1$ and $\lambda_2$ are constants. Then, defining
$J^i{}_\alpha=\epsilon_{\alpha\beta} J^{i\beta}$, the resulting
nontrivial $SL(2)$ duality transformations are
$$
\eqalignno{
\delta P^i{}_\alpha=&\,a P^i{}_\alpha+b J^i{}_\alpha \ ,&(10a)\cr
\delta J^i{}_\alpha=&\,c P^i{}_\alpha-a J^i{}_\alpha\ , &(10b)\cr
\delta\lambda_1=&\, b+2a\lambda_1-c\lambda_1^2+c\lambda_2^2\ , &(10c)\cr
\delta\lambda_2=&\, 2(a-c\lambda_1)\lambda_2 \ . &(10d)\cr}
$$
Note that (10c,d) is the infinitesimal form of a {\it finite}
transformation of the complex field $\lambda=\lambda_1+i\lambda_2$,
of the form $\lambda'={A\lambda+B\over C\lambda+D}$.
The celebrated $R\rightarrow 1/R$ duality transformation corresponds
to $A=D=0$, $B=-C=1$.
We will be seeking the analog of the above $SL(2)$ symmetry in theories
of $p$--branes in $m+p+1$ dimensions.

The dynamical variables describing the $p$--brane are again scalar fields
$x^\mu(\sigma)$, $y^\alpha(\sigma)$ and a worldvolume metric
$\gamma_{ij}(\sigma)$.
Here $\sigma^i\  (i=0,...,p)$ are the worldvolume coordinates.,
$x^\mu$, $\mu=0,...,m-1$ are coordinates on $m$ dimensional space $M$ and
$y^\alpha$, $\alpha=1,...,p+1$ are coordinates on a compact
$(p+1)$-dimensional manifold $N$.
The background fields are as before the metrics $g_{\mu\nu}(x)$ and
$g_{\alpha\beta}(x)$ on $M$ and $N$ respectively
and an antisymmetric tensor field
$b_{\alpha_1\ldots \alpha_{p+1}}(x)
=\lambda_1(x)\epsilon_{\alpha_1\ldots \alpha_{p+1}}$.

The action for the $p$--brane is
$$
\eqalign{
S =\int d^{p+1}\sigma {\cal L}
= \int d^{p+1}\sigma &
\Bigl[-{1\over2}\sqrt{-\gamma}
\left(\gamma^{ij}\p_i x^\mu\p_j x^\nu g_{\mu\nu}
+\gamma^{ij}\partial_iy^\alpha \partial_jy^\beta g_{\alpha\beta}\right)
+{p-1\over2}\sqrt{-\gamma}\cr
&+{1\over(p+1)!}  \epsilon^{i_1\ldots i_{p+1}}
\p_{i_1}y^{\alpha_1}\cdots\p_{i_{p+1}}y^{\alpha_{p+1}}
\lambda_1\epsilon_{\alpha_1\ldots \alpha_{p+1}}\Bigr]\ .\cr}\eqno(11)
$$

The field equations for the variables $x^\mu$ following from the action
(11)  are
$$
\eqalign{
g_{\mu\nu}\p_i\bigl(\sqrt{-\gamma}\gamma^{ij}\p_j x^\nu\bigr)&+
\sqrt{-\gamma}\gamma^{ij}\p_i x^\rho\p_j x^\nu{\mit\Gamma}_{\mu,\rho\nu}
-{1\over2}\sqrt{-\gamma}\gamma^{ij}\p_i y^\alpha\p_j y^\beta\p_\mu
g_{\alpha\beta}\cr
&
+{1\over (p+1)!}\epsilon^{i_1\ldots i_{p+1}}
\p_{i_1}y^{\beta_1}\cdots\p_{i_{p+1}}y^{\beta_{p+1}}
\epsilon_{\beta_1\ldots \beta_{p+1}}\ \p_\mu\lambda_1=\ 0\ .\cr}
\eqno(12)
$$
The field equations for the variables $y^\alpha$ again take the form (4)
with $P^i{}_\alpha$ now defined by
$$
\eqalignno{
P^i{}_\alpha=&{\partial{\cal L}\over\partial\partial_i y^\alpha}=
-\sqrt{-\gamma}\gamma^{ij}\partial_j y^\beta g_{\beta\alpha}
+\lambda_1 J^i{}_\alpha\ , &(13)\cr
J^i{}_\alpha=&{1\over p!}\epsilon^{ij_1\ldots j_p}
\p_{j_1}y^{\beta_1}\cdots\p_{j_p}y^{\beta_p}
\epsilon_{\alpha\beta_1\ldots\beta_p}\ .
&(14)\cr}
$$
{}From the definition (14), the Bianchi identity
$\partial_i J^i{}_\alpha=0$ follows.
We know from ten dimensional supergravity compactified on a
six--torus that under $SL(2)$ the metrics
$g_{\alpha\beta}$ and $g_{\mu\nu}$ rescale. Therefore let us define
$$
g_{\mu\nu}=\lambda_2^K \bar g_{\mu\nu}\ , \qquad\qquad
g_{\alpha\beta}=\lambda_2^L \bar g_{\alpha\beta}\  , \eqno(15)
$$
where $\bar g_{\alpha\beta}$ and $\bar g_{\mu\nu}$ are assumed to be
inert under $SL(2)$,
and ${\rm det}\ \bar g_{\alpha\beta}=1$. Thus $\lambda_2(x)=\left({\rm det}
g_{\alpha\beta}\right)^{1/(p+1)L}$.
In the case $p=5$ it is known from the $SL(2)$ duality symmetry of the
effective field theory limit that $K=-1$ and $L=1/3$ [9].
As will become clear later the most convenient choice for $L$ is $2/(p+1)$,
however, for the sake of generality we
shall keep the values of these powers arbitrary most of the time.

The equation of motion for the worldvolume metric $\gamma$ gives
$$
\gamma_{ij}=\lambda_2^K\p_i x^\mu\p_j x^\nu \bar g_{\mu\nu}+
\lambda_2^L
\p_i y^\alpha\p_j y^\beta \bar g_{\alpha\beta}\ .\eqno(16)
$$
The discussion of strings in $m+2$ dimensions
leads us to postulate the following
form for the infinitesimal duality transformations
$$
\eqalignno{
\delta P^i{}_\alpha=&\,a P^i{}_\alpha+b J^i{}_\alpha \ ,&(17a)\cr
\delta J^i{}_\alpha=&\,c P^i{}_\alpha+d J^i{}_\alpha\ , &(17b)\cr}
$$
where $a,b,c,d$ are constants. As we did in the string case, we have to
show that (13) is invariant under these transformations, combined with
appropriate transformation rules for the background fields $\lambda_1$ and
$\lambda_2$.

Since all relevant quantities have two indices, it is convenient
to use matrix notation. We define matrices $P$ and $J$
with components $P^i{}_\alpha$ and $J^i{}_\alpha$, matrices
$\py$ with components $(\py)^\alpha{}_i=\p_i y^\alpha$ and
$\bar g^{(p+1)}$ with components
$\bar g_{\alpha\beta}$.
{}From the definition (14) we see that the matrix $J$ is directly
related to the inverse of the matrix $\py$ as
$J=\py^{-1}\det\ \py$, and $\det\,(\partial y)=(\det J)^{1/p}$. Therefore we
have the relation
$$
\p y=J^{-1}(\det J)^{1/p}\ . \eqno(18)
$$
This equation allows us to calculate the variation of $\p y$ under the
duality transformations. Using (17b), one finds
$$
\delta\py=\p y\bigg\{-cX+{1\over p}\left(d+c{\rm tr} X\right) \bigg\}
\ , \eqno(19)
$$
where
$$
X=P\cdot J^{-1}\ . \eqno(20)
$$
Note that when $P$ and $J$ are transformed linearly,
$X$ undergoes a fractional linear transformation.
{}From (13) and (16) we find
$$
X=\lambda_1+\lambda_2^L{\gamma^{-1}V \over \sqrt{-\det(\gamma^{-1}V)}}\ ,
\eqno(21) $$
where $V=(\partial y)^T\bar g^{(p+1)}\partial y$.

Following the same steps that led to equation (8), now taking
into account also the variation of $\gamma$, we find after some algebra
that the invariance of (13) under the transformations (17) requires that
$$
\eqalign{
&c X^2\ +\ \left[a-2c\lambda_1
-{1\over p}(d+c\tr X)-\left({p-1\over2}K+L \right)
\lambda_2^{-1}\delta\lambda_2
\right]X
\cr
&+b-{p-1\over p}d\lambda_1+{1\over p}c\lambda_1\tr X
+\left(
{p-1\over2}K+L \right)\lambda_1\lambda_2^{-1}\delta\lambda_2
-\delta\lambda_1=
\cr
=&
\lambda_2^L\gamma^{-1}V\Biggl\{
2c X^2-\left[{2\over p}(d+c\tr X)+2c\lambda_1
+\left(L-K\right)\lambda_2^{-1}\delta\lambda_2 \right] X
\cr
&+{1\over p}c(\tr X)^2-c \tr(X^2)
+\left({1\over p}c\lambda_1+{1\over p}d
+{1\over 2} (L-K) \lambda_2^{-1}\delta\lambda_2\right)\tr X \cr
&\qquad\qquad\qquad\qquad
-{p-1\over p}d\lambda_1
-{1\over 2} (p-1)(L-K)\lambda_1\lambda_2^{-1}\delta\lambda_2
\Biggr\}\ . \cr}\eqno(22)
$$
Since $\gamma^{-1}V$ can be reexpressed in terms of $X$ via Eq. (21),
this is an infinite polynomial equation in the $(p+1)\times(p+1)$
matrix $X$. One is free to determine $\delta\lambda_1$ and
$\delta\lambda_2$ as functions of $a,b,c,d,\lambda_1,\lambda_2$
to satisfy this equation, and also if necessary to put
restrictions on the transformation parameters $a,b,c,d$.
It is important to realize that these transformations have to
be the same for all $X$. In order to prove that this equation
has no solution it would therefore be sufficient to find two
particular matrices $X$ which  give incompatible values
for the variations $\delta\lambda_1$ and $\delta\lambda_2$.

We shall implement this idea by choosing a particular
matrix $X$ and expanding equation (22) around it.
For our background we choose the fields $x^\mu$ and $y^\alpha$
such that $\gamma^{-1}V=\lambda_2^{-L}\eta$.
For example one can have $V_{ij}=\lambda_2^{-L}\delta_{ij}$ and
$\gamma_{ij}=\eta_{ij}$.
Then writing
$$
\gamma^{-1}V=\lambda_2^{-L}(\eta+Y)\eqno(23)
$$
we can expand equation (22) in powers of $Y$.
The zeroth order terms determine the form of the variations
$$
\eqalignno{
\delta\lambda_1=& b+(a-d)\lambda_1
+\left({2\over p}d+
{2\over p+1}{L-K\over L}(a-d)\right)\lambda_2^{(p+1)L/2}\cr
&-c\lambda_1^2
+2c\left({1\over p}-{2\over p+1}{L-K\over L}\right)
\lambda_1\lambda_2^{(p+1)L/2}
+c{3p-2\over p}\lambda_2^{(p+1)L}\ ,&(24a)\cr
\delta\lambda_2=&{2\over (p+1)L}\left((a-d)-2c\lambda_1\right)
\lambda_2^{(p+1)L/2}\ .&(24b)\cr}
$$
At linear order in $Y$ one gets terms proportional to
the matrices ${\bf 1}$, $\eta$, $Y$ and $Y\eta$.
Since these are linearly independent, one can put separately their
coefficients to zero.
In this way one finds
$$
\eqalign{
c=&\, 0\ ,\cr
d=&\, p{K-L\over pK+L}a\ , \cr}\eqno(25)
$$
which inserted in (24) yield
$$
\eqalign{
\delta\lambda_1=& b+{(p+1)L\over pK+L}a\lambda_1\ ,
\cr
\delta\lambda_2=& { 2\over pK+L}a\lambda_2 \ . \cr}\eqno(26)
$$
Remarkably, these transformations give a solution of the full
equation (22). Furthermore, varying the $x^\mu$-equation of motion (12)
under these transformations, we find that its form is preserved.
Therefore, we have a two parameter group of duality transformations
of the $p$--brane.
It is easy to check that the
transformation rules (17) and (26) yield the same commutator algebra.
Denoting the transformations by $a$ and $b$, the only nonvanishing
commutator is $[a,b]=b$.
One can have $d=-a$ by choosing $K={p-1\over 2p}L$.
If we further choose $L=2/(p+1)$, (26) becomes a special case
of (10c,d). In this way
the two parameter group appears to be a subgroup of the expected
group $SL(2)$. Notice that
from a solution with magnetic charge, one can obtain a solution
with magnetic and electric charge.
However, the two parameter group does not contain the important
$R\rightarrow 1/R$ transformations mentioned earlier.

It may be of some interest to consider the special case $m=0$,
namely a $p$-brane propagating in a $p+1$-dimensional space.
This cannot immediately be obtained from the previous discussion
by putting $x^\mu=0$, since the action would be complex.
Instead, we have to assume now that the metric $g_{\alpha\beta}$
is Lorentzian. Then, equation (21) is replaced by
$X=\lambda_1+\lambda_2$, where we have chosen $L=2/(p+1)$.
{}From (17) and (20) one easily finds that
$\delta X=b+(a-d)X-cX^2$.
Therefore it follows that
$\delta(\lambda_1+\lambda_2)=b+(a-d)(\lambda_1+\lambda_2)
-c(\lambda_1+\lambda_2)^2$.
This is the infinitesimal form of a fractional linear transformation
of the real variable $\lambda_1+\lambda_2$ and gives a representation of
the algebra $GL(2)=SL(2)\times R$.
\footnote{$^\dagger$}{Unlike in equation (10c,d) it is not possible to
define the variations of $\lambda_1$ and $\lambda_2$ separately.
If we had kept the metric $g_{\alpha\beta}$ Euclidean we would have
found $X=\lambda_1+i\lambda_2$, leading to separate transformation
rules. However this choice is clearly unphysical since the action
(11) would become complex.}

The subgroup $R$, corresponding to $d=a$, $b=c=0$,
acts trivially on the background fields and
does not mix the field equations with Bianchi identities.
Thus, we can regard this  $R$ as ``trivial''. A similar
trivial factor is present also in the duality group
$GL(1,C)=U(1)\times R$ of a free Maxwell field, see footnote on p. 222
in [14]. Note that the duality group $SL(2)$ of the $p$-brane
in $p+1$ dimensions is the generalization of the nontrivial
$SL(2)$ of the string in $d+2$ dimensions discussed earlier.

There is also an analog of the trivial $SL(2)$.
In the case we are considering, with $L=2/(p+1)$,
Eq. (13) reduces to
$P^i{}_\alpha=\big(\lambda_1+\lambda_2\big) J^i{}_\alpha$,
so it is clear that there
is a larger set of transformations which leave this equation
invariant, without mixing $P$ and $J$, namely
$\delta \lambda_1=0$, $\delta\lambda_2=0$,
$\delta P^i{}_\alpha=\big(R_\alpha{}^\beta
+S^{(\alpha)}\delta_\alpha^\beta \big) P^i{}_\beta$ and  $\delta
J^i{}_\alpha=\big( R_\alpha{}^\beta +S^{(\alpha)}\delta_\alpha^\beta
\big) J^i{}_\beta$ where
$R_\alpha{}^\beta$ are real traceless matrices and $S^{(\alpha)}$ are
the real scaling parameters. These transformations form the group
$SL(p+1,R)\times R^{p+1}$. This group has been discussed in Ref. [13] as a
manifest duality symmetry of the $p$-brane.

The case of $p$-branes in $p+1$ dimensions is very special since
it has only a finite number of degrees of freedom [15] and therefore
is in a certain sense a topological field theory.
Above we have also considered the case of the $p$-brane in $m+n$
dimensions, with $n=p+1$ and a specific choice of background fields.
In Ref. [13] $m$ was taken to be zero, but the dimension $n$ of the
internal space was taken greater than $p+1$, and the backgrounds
were constant.
In particular, in the case $p=2$, $n=4$ it was suggested that there
is an $SL(5,R)$ duality group. However, one should check
the consistency of the transformation rules assigned to all objects.
Specifically, in [13] transformation rules were assigned to the quantities
${\cal F}^{i\alpha}=\sqrt{-\gamma}\gamma^{ij}\partial_j y^\alpha$
and
$\tilde{\cal F}^{i\alpha\beta}=\epsilon^{ijk}\partial_j y^\alpha
\partial_k y^\beta$.
These quantities, however, are related to each other by the relation
${\cal F}^{i\alpha}{\cal F}^{j\beta}\epsilon_{ijk}=
-\gamma_{k\ell}\tilde{\cal F}^{\ell\alpha\beta}$
(we are using here our convention for the Levi-Civita tensor).
The transformation rules given for ${\cal F}^{i\alpha}$ and
$\tilde{\cal F}^{i\alpha\beta}$ are inconsistent with this relation
(the inconsistency arises for the $B^{\alpha\beta\gamma}$
transformations, which are the membrane generalization of the
$C^{\alpha\beta}$ transformations of Eq. (7)).
This, and the results given earlier in this paper, seem to suggest
that the intrinsic nonlinearities of $p$-brane theories will make
it very hard to establish analogs of the ``$C$--symmetries''.

On the other hand, the results of [7] strongly suggest the
existence of an $SL(2,Z)$ duality symmetry of five-branes.
To reconcile this with our results perhaps one should consider
a modification of (17).
It is known that the charges
$Q^J_\alpha=\int d^p\sigma J^0{}_\alpha$
and
$Q^P_\alpha=\int d^p\sigma P^0{}_\alpha$
have to transform under $SL(2)$ as
$\delta Q^P_\alpha=aQ^P_\alpha+bQ^J_\alpha$ and
$\delta Q^J_\alpha=cQ^P_\alpha+dQ^J_\alpha$
[9], but maybe the currents themselves have a slightly different
transformation compatible with this transformation of the charges.
A current algebra argument shows that if the charges transform
like this, the currents have to transform as in (17) up to the
addition of a conserved current whose total charge is zero.
We have not been able to find such a current in this theory.
A more radical departure from our ansatz would be to assume that
$SL(2)$ acts in a nonlocal way, as discussed in [16] in the string
case. Another possible generalization would be to relax the
assumption that all components of the current $J^i{}_\alpha$, namely,
$J^0{}_\alpha, J^1{}_\alpha,...,J^p{}_\alpha$ transform in the same way.
The time components could transform differently from the space
components. This applies also to the components of $P^i{}_\alpha$.
It is natural to consider this idea in the Hamiltonian formalism [17].
Again, this has not led so far to any viable generalization of (17).
Finally, it is possible that only $SL(2,Z)$ is a symmetry of the
five-brane and not $SL(2,R)$. If this is the case, the issue will
have to be settled with methods other than those considered
here, which are based on infinitesimal transformations.

\bigskip\bigskip
\centerline{\bf Acknowledments}
\bigskip

We thank I. Bakas, M. Duff and F. Quevedo for stimulating discussions.
E.S. also thanks
the International Center for Theoretical Physics in Trieste for hospitality.

\vfill\eject

\bigskip
\centerline{\bf References}
\bigskip
\item{1.} K.S. Narain, M.H. Sarmadi and E. Witten, Nucl. Phys. {\bf B279}
         (1987) 367;
\item{} A. Shapere and F. Wilczek, Nucl. Phys. {\bf B320} (1989) 669;
\item{} A. Giveon, E. Rabinovici and G. Veneziano, Nucl. Phys. {\bf B322}
        (1989) 167;
\item{} A. Giveon, N. Malkin and E. Rabinovici, Phys. Lett. {\bf 220} (1989)
         551;
\item{} W. Lerche, D. L\"ust and N.P. Warner, Phys. Lett. {\bf B231} (1989)
417.
\item{2.} M.J. Duff, Phys. Lett. {\bf B173} (1986) 289;
\item{} S. Cecotti, S. Ferrara and L. Girardello, Nucl. Phys. {\bf B308} (1988)
        436;
\item{} J. Molera and B. Ovrut, Phys. Rev. {\bf D40} (1989) 1146;
\item{} M.J. Duff, Nucl. Phys. {\bf B335} (1990) 610;
\item{}  J. Maharana and J.H. Schwarz, Nucl. Phys. {\bf B390} (1993) 3.
\item{3.} E. Cremmer, J. Scherk and S. Ferrara, Phys. Lett. {\bf B74} (1978)
         61;
\item{} M. De Roo, Nucl. Phys. {\bf B255} (1985) 515.
\item{4.} A. Chamseddine, Nucl. Phys. {\bf B185} (1981) 103;
\item{} S. Ferrara, C. Kounnas and M. Porrati, Phys. Lett. {\bf B181}
         (1986) 263;
\item{} M. Terentev, Sov. J. Nucl. Phys. {\bf 49} (1989)713;
\item{} S.F. Hassan and A. Sen, Nucl. Phys. {\bf B375} (1992) 103.
\item{5.} A. Giveon, M. Porrati and E. Rabinovici, preprint RI-1-94
          (hep-th/9401139).
\item{6.} A. Font, L. Ib\'a\~nez, D. L\"ust and F. Quevedo, Phys. Lett. {\bf
          B249} (1990) 35;
\item{} S.J. Rey, Phys. Rev. {\bf D43} (1991) 526;
\item{} A. Shapere, S. Trivedi and F. Wilczek, Mod. Phys. Lett. {\bf A6}
          (1991) 2667;
\item{} A. Sen, Nucl. Phys. {\bf B404} (1993) 109;
\item{} J. Schwarz, preprint, CALT-68-1815 (hep-th/9209125).
\item{7.} A. Sen, Nucl. Phys. {\bf B388} (1992) 457;
           Phys. Lett. {\bf B303} (1993) 22;
           Mod. Phys. Lett. {\bf A8} (1993) 2023;
 \item{} J.H. Schwarz and A. Sen, Phys. Lett. {\bf 312} (1993) 105;
\item{} T. Ortin, Phys. Rev. {\bf D47} (1993) 3136;
\item{} M.J. Duff and R. Khuri, preprint, CTP-TAMU-17/93
\item{8.}  A. Tseytlin, Phys. Lett. {\bf B232} (1990) 163; Nucl. Phys. {\bf
           B350} (1991) 395;
\item{}J.H. Schwarz and A. Sen, Nucl. Phys. {\bf B411} (1994) 35;
\item{} J.H. Schwarz, preprint, CALT-68-1879 (hep-th/9307121).
\item{9.}   A. Sen, preprint, TIFR/TH/94-03 (hep-th/9402002).
\item{10.} M.J. Duff, Class. Quant. Grav. {\bf 5} (1988) 189;
\item{} A. Strominger, Nucl. Phys. {\bf B343} (1990) 167;
\item{} M.J. Duff and J.X. Lu, Nucl. Phys. {\bf B354} (1991) 141; Phys. Rev.
         Lett. {\bf 66} (1991) 1402; Class. Quant. Grav. {\bf 9} (1991) 1;
\item{} C.G. Callan, J.A. Harvey and A. Strominger, Nucl. Phys. {\bf B359}
       (1991) 611; Nucl. Phys. {\bf B367} (1991) 60;
\item{} M.J. Duff, R. Khuri and J.X. Lu, Nucl. Phys. {\bf B377} (1992) 281;
\item{} J. Dixon, M.J. Duff and J. Plefka, Phys. rev. Lett. {\bf 69} (1992)
        3009.
\item{11.} J.H. Schwarz and A. Sen, Phys. Lett. {\bf B312} (1993) 105;
\item{} J.H. Schwarz and A. Sen, Phys. Lett. {\bf B312} (1993) 105;
\item{}   P. Binetr\'uy, Phys. Lett. {\bf B315} (1993) 80.
\item{12.} E. Bergshoeff, E. Sezgin and P.K. Townsend, Phys. Lett. {\bf 189B}
          (1987) 75.
\item{13.} M.J. Duff and J.X. Lu, Nucl. Phys. {\bf B347} (1990) 394.
\item{14.} M.K. Gaillard and B. Zumino, Nucl. Phys. {\bf B193} (1981) 221.
\item{15.} K. Fujikawa, Phys. Lett. {\bf B 213} (1988) 425;\hfil\break
R. Floreanini and R. Percacci, Mod. Phys. Lett. {\bf A 5} (1990) 2247.
\item{16.} I. Bakas, preprint CERN-TH.7144/94 (hep-th/9402016).
\item{17.} E. Sezgin, talk given at the G\"ursey Memorial Conference I
on Strings and Symmetries, Istanbul, Turkey, June 6-10, 1994.
\end